\documentclass[aip,amsmath,amssymb,reprint]{revtex4-1}
\usepackage{color}
\usepackage{graphicx}
\usepackage{dcolumn}
\usepackage{bm}
\usepackage[utf8]{inputenc}
\usepackage[T1]{fontenc}
\usepackage{mathptmx}
\usepackage{units}
\usepackage{textcomp}
\usepackage{ulem}
\usepackage{braket}
\usepackage{hyperref}

\begin{document}

\title{Dynamical low-noise microwave source for cold-atom experiments}

\author{Bernd Meyer-Hoppe}
\email[]{The author to whom correspondence may be addressed: meyer-hoppe@iqo.uni-hannover.de}
\author{Maximilian Baron}
\author{Christophe Cassens}
\author{Fabian Anders}
\author{Alexander Idel}
\author{Jan Peise}
\author{Carsten Klempt}
\affiliation{Leibniz Universit\"at Hannover, Institut für Quantenoptik, Welfengarten 1, D-30167 Hannover, Germany}

\date{\today}

\begin{abstract}
The generation and manipulation of ultracold atomic ensembles in the quantum regime require the application of dynamically controllable microwave fields with ultra-low noise performance.
Here, we present a low-phase-noise microwave source with two independently controllable output paths.
Both paths generate frequencies in the range of $6.835\,$GHz $\pm$ $25\,$MHz for hyperfine transitions in $^{87}$Rb.
The presented microwave source combines two commercially available frequency synthesizers: an ultra-low-noise oscillator at $7$~GHz and a direct digital synthesizer for radiofrequencies.
We demonstrate a low integrated phase noise of $480\,${\textmu}rad in the range of $10\,$Hz to $100\,$kHz and fast updates of frequency, amplitude and phase in sub-{\textmu}s time scales.
The highly dynamic control enables the generation of shaped pulse forms and the deployment of composite pulses to suppress the influence of various noise sources.
\end{abstract}

\maketitle

\section{Introduction} \label{intro}
Atom interferometers belong to today's most precise sensors with broad applications for navigation, geodesy, time keeping as well as fundamental research.
An atom interferometric measurement relies on the determination of the phase difference between two atomic states.
In many cases, these atomic states are hyperfine levels in alkali atoms.
For magnetic field sensing or frequency measurements, the hyperfine transitions can be directly driven with microwave radiation.
For inertial sensors, such as gravimeters, accelerometers or gyroscopes, the hyperfine transitions can be driven by a two-photon optical transition with a microwave frequency detuning.
In analogy to optical interferometers, microwave pulses driving these transitions act as beam splitters or mirrors between the atomic states.
Phase noise of the microwave is directly converted into fluctuations of the measured interferometer phase.
The rapidly increasing sensitivity in the field of atom interferometry poses increasing demands on the noise characteristics of the employed microwave sources.
This is specifically the case for interferometric experiments with squeezed atomic samples~\cite{Pezze2018}, which are not restricted by the Standard Quantum Limit (SQL) and are thus even more sensitive to the microwave's phase and intensity noise.

\begin{figure*}
\includegraphics[width=1\textwidth]{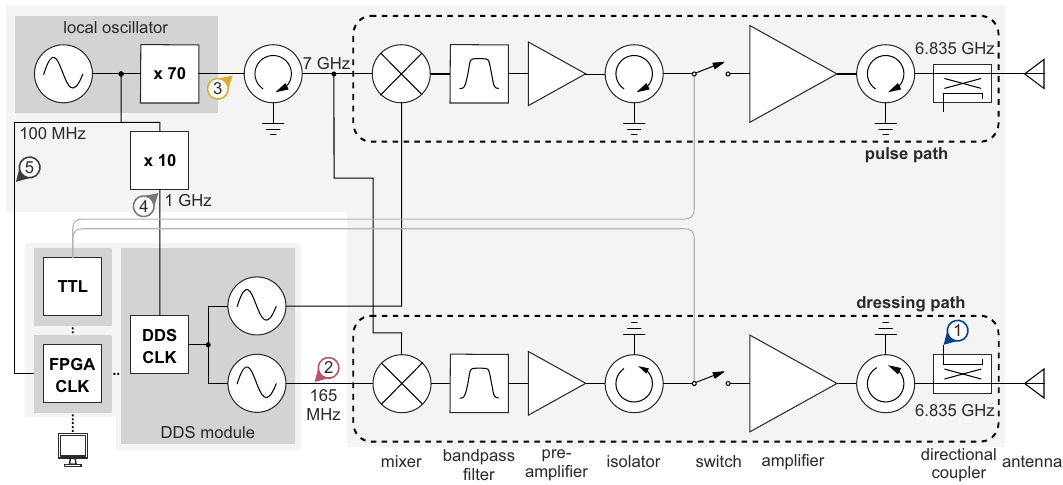}%
\caption{\label{setup}Schematic of the microwave source.
The direct digital synthesizer (DDS) module generates two independently controllable radiofrequency signals.
Each of these signals is mixed with a $7\,$GHz frequency coming from an ultra-low-phase-noise oscillator with an isolator protecting the oscillator from back reflections.
The local oscillator's $100\,$MHz output is split into two paths:
One path acts as a reference for the FPGA's clock while the other one is multiplied by 10 and subsequently delivers the frequency reference for the DDS module, thereby ensuring a steady phase relation between the oscillator and the DDS.
In this way, two controllable microwave frequency paths are established.
The pulse path consists of a pre-amplifier and a narrow bandpass filter to generate the optimal input for the high power amplifier.
Moreover, isolators prevent back reflections from a switch and the antenna.
A fraction of the amplified signal, branched off with a directional coupler, enables monitoring the microwave power.
In the dressing path, the same basic schematic is used, except for a different power amplifier.
The large part with light gray background is mounted on a water-cooled base plate and a photo of it is shown in Fig.~\ref{pic}.
The numbers and colors correspond to the measured phase noise in Fig. \ref{budget}.}%
\end{figure*}

In our experiments, we create entangled many-particle states in Bose-Einstein condensates by spin-changing collisions~\cite{Lucke2011,Lucke2014,Kruse2016,Lange2018}.
The microwave source serves four purposes: The initial preparation of all atoms in a specific hyperfine/Zeeman level, the dressing of Zeeman levels to activate spin-changing collisions, the manipulation of the states in an interferometric protocol, and the final rotation of the state for detection and state analysis.
These tasks pose the following requirements on the design of the microwave source.

\textit{Variable frequency.} For the preparation of the atoms, a sequence of microwave pulses with variable frequencies is desired.

\textit{Low phase noise.} The high-fidelity manipulation of the atomic states demands a microwave with small phase fluctuations, as the phase serves as the reference for all atomic measurements~\cite{Chen2012a}.
In our atom interferometric applications, the fluctuation of the microwave's phase during the interferometric cycle is directly converted into fluctuations of the quantity of interest.
For the future, we would like to operate the interferometers with microwave pulses of $10\,$\textmu{}s length and free evolution times of several tens of milliseconds, such that a fluctuating phase is relevant from $10\,$Hz up to $100\,$kHz.
We aim at an interferometric phase resolution close to the Heisenberg limit $\Delta\Phi\gtrsim 1/N$, with the number of employed particles $N$.
The generation of entangled states is currently performed with $N=10^4$ atoms.
A Heisenberg-limited resolution for such atom numbers of course requires the suppression of more technical noise sources.
For example, we were able to improve the so far limiting detection noise to single-atom resolution for several hundred atoms~\cite{Hueper2020}.
To ensure that the microwave phase noise is not limiting, it must be suppressed to a value below $\Delta\Phi < 2\pi / N \approx 600$\textmu{}rad.

\textit{Low intensity noise.} For tomographic measurements of the entangled states, the microwave should perform stable couplings between the employed spin levels. To resolve the expected patterns with single-particle resolution, the microwave must offer spin rotations with a relative intensity stabilization $\Delta I / I < 2/N \approx 0.02\%$ that is stable for a typical measurement time on the order of one hour.

\textit{Fast dynamical adjustments.} The spin preparation benefits from short microwave pulses that are executed closely after each other.
Short pulses bear the advantage of a low sensitivity to small microwave detunings due to an increasingly broad peak in Fourier space.
This is, however, not sufficient for an advantageous spectral distribution.
Short pulses can lead to many side peaks in the Fourier spectrum if they are simple box pulses.
This may lead to an unwanted population of neighboring levels.
The entire width of the spectrum can be drastically reduced by applying tailored pulse shapes instead of rectangular pulses.
Pulse shaping requires a microwave amplitude modulation below the microsecond scale and the short pulse durations can only be realized by a source with sufficient microwave power.
Furthermore, some interferometry and state tomography schemes are based on a fast and reproducible adjustment of the microwave's phase.
The combined modulation of frequency, phase and intensity enables the application of composite pulses~\cite{Dunning2014}, which can be designed to suppress the sensitivity to various technical noise sources.

\textit{Intensity-stable dressing field.} Finally, we wish to apply an independent microwave dressing field during the experimental sequence to achieve precise control of individual energy levels.
This application is particularly sensitive to fluctuations of the microwave field's intensity because the desired energy shift depends linearly on the power of the microwave signal.
Hence, a stable microwave power is desired for quasi-adiabatic state preparation by ramped microwave dressing, as we use it for the generation of highly entangled twin-Fock states~\cite{Luo2017,Anders2021}.
Typically, for a $1\,$G magnetic field, we shift the $\ket{F=1,m_F=0}$ level via microwave dressing by $72\,$Hz.
For high-accuracy adiabatic generation, we require maximal frequency fluctuations of $10\,$mHz over typical ramping times of a few seconds, corresponding to microwave intensity fluctuations of $10\,$mHz$/72\,$Hz$=0.014$\%.
In summary, our experiments require a microwave source with amplitude, frequency and phase modulation in the microsecond range, low phase fluctuations and an independent, stable microwave dressing field.

There exist commercial frequency generators in the microwave regime that allow for a dynamical adjustment of frequency and amplitude.
However, these systems typically lack an independent adjustment of the phase or suffer from slow update speeds in the millisecond range.
There are important scientific developments towards microwave sources with minimal phase fluctuations, which serve as local oscillators for microwave clocks~\cite{Boudot2009a,Lipphardt2009,Fortier2012,Francois2015,AbdelHafiz2017}.
However, for these applications, a fast dynamical adjustment is not intended.

Publications reporting on low-phase-noise microwave sources that are designed for a fast dynamical adjustment of frequency, phase and amplitude are rare.
Chen et al.~\cite{Chen2012b} present a microwave source for spin-squeezing experiments in $^{87}$Rb based on a self-built nonlinear transmission line that is phase-locked to a $10\,$MHz atomic clock.
Controllability of parameters in the range of $24\,${\textmu}s to $32\,${\textmu}s is achieved by a single-sideband modulator and a direct digital synthesizer (DDS).
The obtained integrated phase noise amounts to $3\,$mrad in the $10\,$Hz to $100\,$kHz bandwidth.
Morgenstern et al.~\cite{Morgenstern2020} demonstrated a microwave source at $1.8\,$GHz, controlled by a field programmable gate array (FPGA), to realize a microwave dressing in sodium Bose-Einstein condensates.
The source features parameter changes within $4\,${\textmu}s.
The phase stability is not specified, as it is not critical for their application.

Here, we present a microwave source with low integrated phase noise of $480\,${\textmu}rad in the $10\,$Hz to $100\,$kHz bandwidth.
The microwave frequency is derived from a commercial $7\,$GHz source and a DDS, which controls frequency, amplitude and phase with high accuracy.
An interplay of components from the open-source hardware family \textit{Sinara} \footnote{Information about the Sinara ecosystem can be found under https://sinara-hw.github.io.} and the control system \textit{ARTIQ}~\cite{artiq2021} enables single-parameter updates within $712\,$ns.
The setup does not comprise self-built components and fits into a 19 inch rack.
The source consists of two independent DDS-controlled channels, which can be applied for two-photon transitions or for a background dressing field.
The described microwave source thus offers the ability to control atomic spins with highest fidelity on short time scales, and the simultaneous application of a dressing field.

The article is organized as follows.
Section \ref{system design} covers the system design and the installed components.
In Section \ref{characterization}, we characterize the system regarding phase noise and dynamic features.
Finally, we summarize our findings and outline the future application of the system in our spin-squeezing experiments.

\begin{figure}
\includegraphics[width=0.45\textwidth]{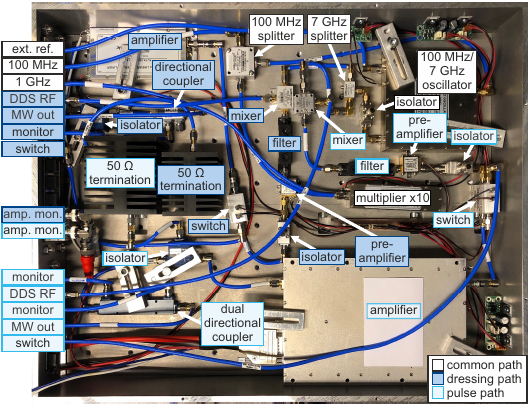}%
\caption{\label{pic}Picture of the developed microwave source without the \textit{Sinara} modules with labels for external connections and important components. Coloring identifies dressing path (dark blue), pulse path (light blue), and common path (white).
This part of the microwave source is mounted on a water-cooled base plate.
It is completed by a 19 inch rack containing an FPGA controller, a DDS module and a digital input/output module (not shown here).
A list of components is provided online~\cite{mwcomponents2023}.
}%
\end{figure}

\section{System Design\label{system design}}
Figure~\ref{setup} shows a schematic of the final design of our microwave source and Fig.~\ref{pic} the corresponding picture.
The basic idea is to combine the low phase noise of a multiplied crystal oscillator with the dynamic properties of an FPGA-controlled DDS.
Our design presents an easy-to-assemble setup with commercial components~\cite{mwcomponents2023} and very low phase noise and our implementation opens up highly dynamical features.

An ultra-low-phase-noise oscillator \textit{Wenzel MXO-PLD} is used as local oscillator.
It provides a stable microwave frequency at $7\,$GHz through integrated multiplication of a fundamental $100\,$MHz oscillation, which is accessible through an additional output.
The oscillator also has the option of an external reference, which is currently not used in our setup.

For the dynamic parameter change of the radiofrequency component, we employ the commercially available DDS module \textit{Urukul}~\cite{Kasprowicz2022}.
This module contains four \textit{Analog Devices AD9910} \footnote{The product page of the \textit{AD9910} DDS can be found under https://www.analog.com/en/products/ad9910.html?doc=AD9910.pdf.} DDS chips and associated radiofrequency components for adequate filtering, variable attenuation and switching.
Control over the DDS module is achieved by the FPGA module \textit{Kasli}. A digital I/O module \textit{DIO\_BNC} offers the possibility for triggered execution and controlling additional components such as microwave switches.
All three modules are part of the \textit{Sinara} ecosystem and can be bought commercially as a preconfigured device.
The DDS module features four independently controllable radiofrequency channels with an adjustable frequency of up to $400 \,$MHz.
All channels share a common reference of $1\,$GHz, given by the local oscillator's $100\,$MHz multiplied by 10.
The $1\,$GHz frequency reference allows for direct clocking of the DDS channels.
Compared to the case of indirect clocking, where the $1\,$GHz clock is generated by internally multiplying the local oscillator's $100\,$MHz in the DDS module, we measured an improved phase noise performance of $100\,${\textmu}rad.
For the microwave source, two of these DDS outputs operate the two individual microwave output stages.
In addition to frequency setting, the amplitude and also the phase offset for absolute phase control can be defined (Tab.~\ref{param}).

\begin{table}
  \centering
  \begin{tabular}{p{2.3cm}p{1.3cm}p{2.3cm}p{2cm}}
    \hline\hline
    Parameter & Bits & Range & Resolution\\
    \hline
    Frequency & 32 & $0-400\,$MHz & $230\,$mHz\\
    Amplitude & 14 & $0-100\,$\% & $0.006\,$\%\\
    Phase offset & 16 & $0 - 2\pi\,$rad & $96\,${\textmu}rad\\
    \hline\hline
  \end{tabular}
  \caption{\label{param}Specification of the adjustable parameters of the DDS.
  The frequency range is given such that the attenuation due to the reconstruction filter is maximally $3\,$dB.
  For a maximum attenuation of $1\,$dB, the upper frequency bound is $320\,$MHz.}
\end{table}
For our $^{87}$Rb atoms, we have to apply frequencies at ${f_{Rb}=6.835\,\textrm{GHz}}$ with a tuning range of $\pm 5\,$MHz to account for the employed magnetic fields.
In both output channels, the frequency is reached by mixing a shared local oscillator (LO) signal at $f_{LO}=7\,$GHz  with an individual DDS frequency of $f_{DDS}=165 \pm 5\,$MHz.
The lower sideband after the mixing presents the desired resonant frequency at $f_{Rb}=f_{LO}-f_{DDS}$.
The two output stages for pulses and dressing include final amplifiers to obtain powers of $40\,$W and $10\,$W, respectively.

For the mixing process, a \textit{Marki Microwave SSB-0618} single sideband mixer and a \textit{Mini Circuits ZMX-8GLH} mixer are used for the pulse path and the dressing path, respectively.
The single sideband mixer in the more critical channel is used to lower the unwanted peaks at the local oscillator frequency $f_{LO}$ and the upper sideband at $f_{LO}+f_{DDS}$.
A narrow \textit{Wainwright WBCQV3} bandpass filter with a passband of 20 MHz and ${50\,\textrm{dB}}$ attenuation for frequencies of $\pm 250\,$MHz from the center frequency enables the reduction of the LO and the upper sideband level in both channels.

Directly after mixing, the pulse and dressing paths measure maximally $-7.5\,$dBm and $-6.5\,$dBm, respectively.
Low-noise pre-amplifiers \textit{Mini Circuits ZX60-83LN-S+} raise the signal to the desired $0\,$dBm input for the high-power amplifiers.

Because the internal power switch of the DDS channel only offers a maximal attenuation of $31.5\,$dB, the source is equipped with additional power switches \textit{Mini Circuits ZFSWA2-63DR+} with $25\,$ns rise/fall time and an attenuation of $40\,$dB.
The employed switches -- designed for frequencies up to $6\,$GHz -- are a cost-effective solution, coming with disadvantages of larger insertion loss and reflections.
The insertion loss is compensated by the pre-amplifiers.
Isolators prevent unwanted back reflections in general, and are installed in front of the switches and the antenna, as well as behind the local oscillator.

For the amplification of the pulse path, a water-cooled $40\,$W amplifier \textit{Microwave Amps AM43} is employed.
The amplifier in the dressing path is a \textit{Kuhne KU PA 640720-10 A} with an output power of $10\,$W.
The pulse path contains a $-30\,$dB dual directional coupler \textit{RF-Lambda RFDDC2G8G30} for monitoring the output power and the power that is reflected from the antenna.
A $-20\,$dB directional coupler \textit{MCLI C39-20} after the dressing amplifier allows for monitoring the dressing power.
For addressing the atoms, the microwave frequency is coupled to free space by an impedance-matched, open-ended waveguide.
In the following section, the performance of the described microwave source is evaluated.

\section{Characterization\label{characterization}}
\subsection{Spectral properties}
\begin{figure}
\includegraphics[width=0.45\textwidth]{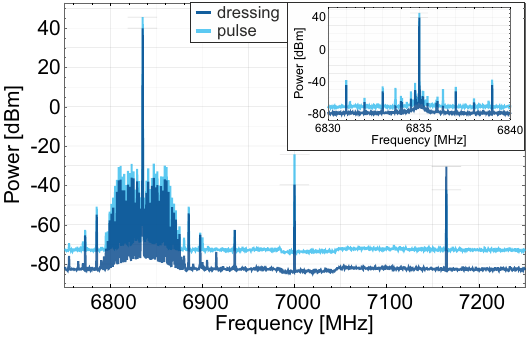}
\caption{\label{spec}
Spectrum of the microwave source for the dressing path (dark blue) and the pulse path (light blue), where the measured power in dBm is plotted as a function of the microwave frequency.
The main difference of the two paths is the maximum power and the suppression of the LO frequency as well as the upper sideband.
In the inset, the relevant frequency range is presented in more detail.
Both spectra have been measured with a resolution bandwidth of $10\,$Hz.}
\end{figure}

First of all, we examine the spectrum of the microwave source to quantify the suppression of disturbing frequencies, that could drive unwanted transitions, resulting in decreased efficiency of the state preparation.
Figure~\ref{spec} shows the spectrum of our microwave source for both the dressing path and the pulse path with a chosen output frequency of $6.835\,$GHz.
The spectrum was recorded with a \textit{Rohde\&Schwarz FSWP8} phase noise and spectrum analyzer.

\begin{figure}
\includegraphics[width=0.45\textwidth]{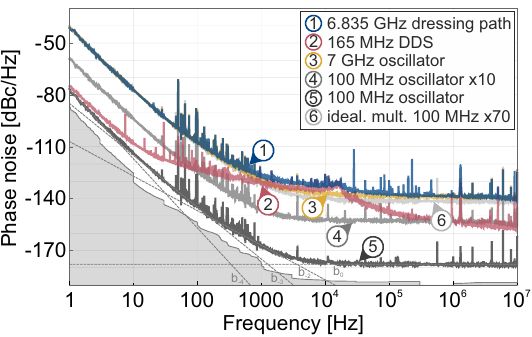}%
\caption{\label{budget}
Phase noise $\mathcal{L}(f)$ at different points of the microwave source plotted against the offset frequency from the carrier.
The color of the lines and the numbers correspond to the colored dots and numbers in Fig.~\ref{setup}.
Because the phase noise of the dressing path and the pulse path is very similar, only the dressing path's phase noise is plotted (1, dark blue).
The output paths' phase noise is dominated by the $165\,$MHz DDS (2, red) only for a small frequency range around $16\,$kHz and by the local oscillator (3, yellow) for all other offset frequencies.
Three gray curves represent the local oscillator's $100\,$MHz (5), the phase noise of the multiplied LO frequency at $1\,$GHz (4) and the ideally multiplied $100\,$MHz by $70$ (6).
For the $100\,$MHz oscillator curve, four dashed lines indicate a parametric estimation of the phase noise, $\mathcal{L}(f)=\sum_{i=-4}^0{b_if^i}$, with coefficients $b_0=-178\,$dBc/Hz, $b_{-2}=-107\,$dBc/Hz, $b_{-3}=-85\,$dBc/Hz and $b_{-4}=-77\,$dBc/Hz.
These data allow for a further characterization of the employed oscillator via 'oscillator hacking'~\cite{rubiola2009}.
The sensitivity limit of the measurement for $100\,$MHz is shown as a gray area.
Measurement parameters are a resolution bandwidth of $1\,$\% and a cross-correlation factor of $1000$.
}%
\end{figure}

The unwanted frequency contributions close to the main peak are caused by the DDS, while the two peaks at $7\,$GHz and $7.165\,$GHz are residuals of the local oscillator frequency and the upper sideband of the mixing process.
Both of these peaks are suppressed by the narrow bandpass filter:
For the pulse path, the local oscillator and the upper sideband peak are suppressed by $69.8\,$dB and $87.9\,$dB, respectively.
For the dressing path, the suppression is $79.5\,$dB and $70.3\,$dB due to the different choice for the frequency mixer.

Because of the bandpass filters, the peak power slightly varies with the frequency of the microwave source by about $\pm 0.2\,$dB in the relevant frequency range of $6.835\,$GHz $\pm 5\,$MHz and $\pm 1\,$dB for $6.835\,$GHz $\pm 25\,$MHz.
The small effect in the relevant range can be calibrated and the power can be adjusted by the \textit{ARTIQ} software.

Furthermore, we evaluate the phase noise $\mathcal{L}(f)=\tfrac{1}{2} S_{\varphi}(f)$, where $S_{\varphi}(f)$ is the one-sided spectral density of phase fluctuations and $f$ is the offset frequency from the carrier~\cite{FerrePikal2009}.
A low phase noise is desired for a high-fidelity state preparation and manipulation (Section \ref{intro}).
Figure~\ref{budget} shows the phase noise contributions for the microwave source, measured for an output frequency of $6.835\,$GHz with the \textit{Rohde\&Schwarz FSWP8}.
The colors of the lines correspond to different measurement points in the system, as visualized in Fig.~\ref{setup}.

The phase noise of the two microwave output signals is very similar, although they follow different amplification paths.
The overall phase noise is limited by the DDS for frequencies from $2\,$kHz up to $20\,$kHz and by the local oscillator for all other frequencies.
It is below $-120\,$dBc/Hz above $350\,$Hz and below $-130\,$dBc/Hz for frequencies greater than $2.5\,$kHz, and approaches a floor of $-140\,$dBc/Hz at high frequencies.
The small peak at $16\,$kHz stems from residual DDS noise~\cite{Calosso2020} and may originate from power supply noise.
For both paths, we obtain a favorably small integrated phase noise of ${\Delta \Phi = 480\,\textrm{{\textmu}rad}}$ in the important range of $10\,$Hz to $100\,$kHz (Section \ref{intro}), which provides the central figure of merit of the constructed microwave source.

With a reference connected to the local oscillator, the long-term stability could be improved and therefore the phase noise for small frequency offsets could be decreased.

\begin{figure}
\includegraphics[width=0.45\textwidth]{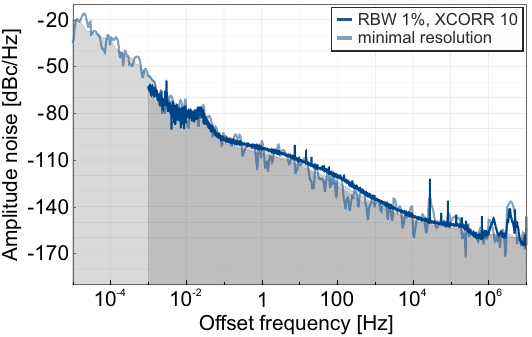}%
\caption{\label{AMbudget}
Amplitude noise $S_{a}(f)$ of the 6.835 GHz dressing path plotted as a function of the offset frequency from the carrier.
The solid curve represents a measurement with a resolution bandwidth of $1\,$\% and a cross-correlation factor of $10$.
The transparent curve corresponds to a minimal resolution setting of a single measurement per half-decade to reach very low offset frequencies.
The sensitivity limits of the measurement are drawn in gray.}%
\end{figure}

The amplitude (AM) noise $S_{a}(f)$ of the presented microwave source is also measured and shown in Fig.~\ref{AMbudget}.
Because the amplitude noise is important also for very low offset frequencies and measurements in that range have a very long duration, high sensitivity limits had to be accepted as a consequence.
Hence, the amplitude noise of the microwave source can only be given as an upper bound.
For the bandwidth of $278\,${\textmu}Hz (measurement duration of one hour) to $100\,$kHz, it evaluates to an integrated AM noise of $0.015\,$\% and therefore an intensity noise of $\Delta I / I = 0.03\,$\%.
It is thus in the desired order of magnitude to achieve Heisenberg-limited resolution.

\subsection{Dynamic features}
The dynamics of the microwave source is controlled by the \textit{ARTIQ} system which is connected to a computer via an optical fiber and a LAN switch.
The commands are implemented on the computer in the \textit{Python} programming language but compiled and executed on the FPGA hardware, thereby allowing for a minimal timing resolution of $4\,$ns.
The code generating our experiments is available online\footnote{The gitlab repository of the control program is avaiable under https://gitlab.projekt.uni-hannover.de/iqo-artiq/artiq/public/arbitrary-pulses-spinor-bec}.

\begin{figure}
\includegraphics[width=0.45\textwidth]{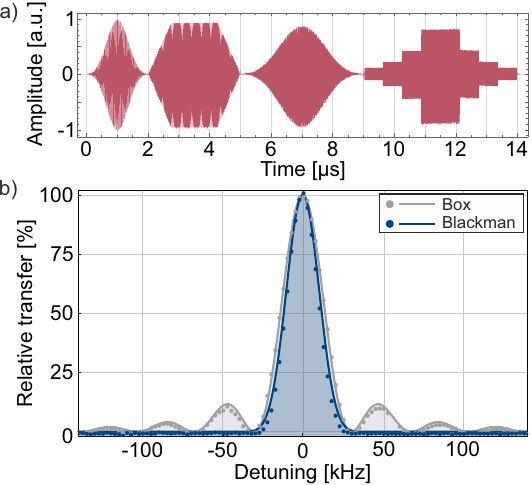}%
\caption{\label{shapes}
(a) Measurement of different amplitude-shaped pulses using the DDS's RAM mode.
The four pulses consist of different shapes (Blackman, linear with plateau, Blackman, staircase), frequencies ($50\,$MHz, $60\,$MHz, $70\,$MHz, $80\,$MHz) and durations ($2\,$\textmu{}s, $3\,$\textmu{}s, $4\,$\textmu{}s, $5\,$\textmu{}s).
Our implementation takes care of the optimal temporal resolution for a given pulse sequence regarding the limited step number of 1024.
The different maximum amplitudes are due to the bandwidth of our oscilloscope.
(b) Comparison of the atomic transfer resulting from a box pulse and a Blackman pulse where the microwave frequency is varied.
The $\pi$ pulse durations are $t_{\pi,\textrm{box}}=28.6\,${\textmu}s and $t_{\pi,\textrm{bm}}=69.3\,${\textmu}s for the box and Blackman pulse, respectively.
Lines represent theoretical simulations with only the resonance frequency as free parameter and dots experimental data.
The resulting resonance frequency of the clock transition is $f_\textrm{res}=6\,834\,683\,$kHz.
}%
\end{figure}

The \textit{AD9910} DDS chip in the \textit{Urukul} module also provides a $1024 x 32\,$ bit random-access memory (RAM) to retrieve and output different waveforms that are designed by the user.
It is possible to program frequency, amplitude, phase or polar (amplitude and phase) nonlinear ramps with a step width of $4\,$ns into one out of eight profiles.
For our purposes, time-dependent amplitude functions are used to reduce the effective Fourier width (Fig.~\ref{shapes}) and the coupling to unwanted transitions.
For the dressing scenario, adiabatic power ramps ensure that the system remains in the desired dressed state.

\begin{figure}
\includegraphics[width=0.45\textwidth]{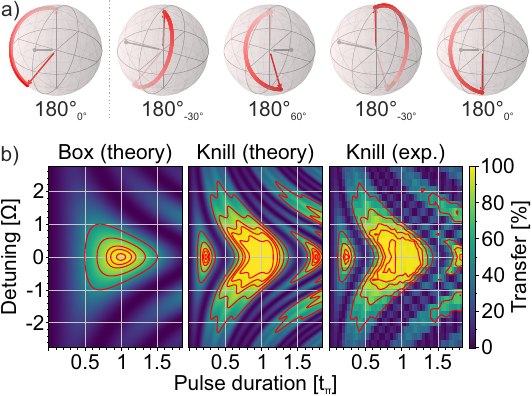}
\caption{\label{knill}
(a) Bloch sphere representation of a \textit{Knill} sequence.
Red arrows indicate state trajectories (from transparent to opaque) and the gray line corresponds to the rotation axis.
Rotations are shown with 90\% $\pi$ pulse duration and a frequency detuning of half a Rabi frequency to present the noise cancelling effect.
The first Bloch sphere shows a box-shaped $\pi$ pulse.
A composite \textit{Knill} pulse consists of four additional $\pi$ pulses around different axes as indicated below the spheres.
(b) Compared to a box pulse, the \textit{Knill} sequence shows a reduced sensitivity to frequency detuning (in multiples of the resonant Rabi frequency $\Omega=27\,$kHz) and pulse duration (in units of the $\pi$ pulse duration $t_{\pi}=115\,$\textmu{}s).
Red lines indicate transfer of \{99, 95, 90, 75, 50\}\%.
Our experimental results resemble the theory very well.
}%
\end{figure}

\begin{figure}
\includegraphics[width=0.45\textwidth]{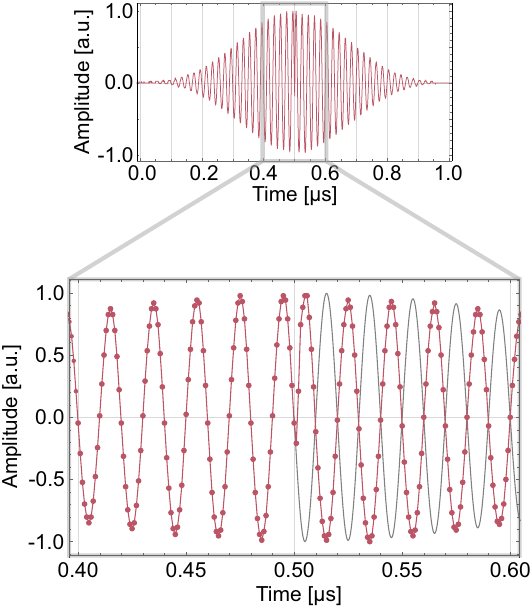}%
\caption{\label{phase change}
Phase control of the radiofrequency source.
At the top, a Blackman-shaped pulse is shown.
The pulse was implemented in a way that the phase is changed in the middle of the pulse.
A detailed view is shown at the bottom.
The dots correspond to the measured data points, the red lines are a guide to the eye.
The grey line indicates the extrapolation of the first half of the pulse, confirming the phase change of half a turn.}%
\end{figure}

The phase of the microwave signal is deterministically referenced to the local oscillator phase in a coherent way.
This allows for composite pulse techniques, where one single pulse is replaced by a sequence of pulses with different phases, amplitudes and durations to reduce the impact of various noise sources~\cite{Dunning2014,Levitt1986}.
Figure~\ref{knill} shows a so-called \textit{Knill} sequence~\cite{ryan2010} that reduces sensitivity to both intensity and frequency fluctuations.
Besides the phase setting, it is advantageous for this technique to execute the pulses in quick succession to avoid unwanted fluctuations during the dead times.
Updates of parameters like frequency or phase require $712\,$ns in our implementation. 
This corresponds to a minimal pulse duration of a few {\textmu}s when two totally different pulses should be executed in direct succession.
Start and end times of pulses on different DDSs have to be equal or at least differ by $424\,$ns if they are on the same \textit{Urukul} module.
Two separate modules could, however, be operated completely independent without this delay.
Figure~\ref{phase change} finally shows an amplitude-shaped pulse where the phase of the radiofrequency is altered halfway using one common profile instead of two distinguished pulses.
This example demonstrates the highly dynamical phase control.

\section{Conclusion}
In summary, we have presented a microwave source with sub-{\textmu}s update speed and adjustable phase.
Composite pulses and nonlinear ramps can be delivered while maintaining a low integrated phase noise of $480\,${\textmu}rad.
The experimental sequences are implemented on a computer with \textit{Python} and the \textit{ARTIQ} software, allowing for a nanosecond timing resolution.
The microwave source is based on commercially available components and adaptable to other atomic species than $^{87}$Rb.
The design with two independently controllable paths opens up various possibilities such as the addressing of two-photon processes and the Rabi coupling to a dressed state.\\

\begin{acknowledgments}
We thank \'{E}. Wodey for his support regarding the ARTIQ implementation.
We acknowledge financial support from the Deutsche Forschungsgemeinschaft (DFG, German Research Foundation) -- Project-ID 274200144 -- SFB 1227 DQ-mat within the project A02, under Germany’s Excellence Strategy -- EXC-2123 QuantumFrontiers -- 390837967 and from the European Union through the QuantERA grant 18-QUAN-0012-01 (CEBBEC). 
F.A. and B.M.-H. acknowledge support from the Hannover School for Nanotechnology (HSN).
B.M-H. acknowledges the use of the QuTiP~\cite{Johansson2013} Python toolbox for Bloch sphere illustrations.
\end{acknowledgments}

\section*{Conflict of Interest Statement}
The authors have no conflicts to disclose.

\section*{Author contributions}
\textbf{Bernd Meyer-Hoppe}: Conceptualization (lead); Investigation (lead); Software (supporting); Visualization (lead); Writing - Original Draft Preparation (equal); Writing - Review \& Editing (equal). 
\textbf{Maximilian Baron}: Investigation (supporting); Software (lead); Visualization (supporting); Writing – Review \& Editing (equal). 
\textbf{Christophe Cassens}: Investigation (supporting); Writing – Review \& Editing (equal). 
\textbf{Fabian Anders}: Conceptualization (supporting); Supervision (supporting); Writing – Review \& Editing (equal). 
\textbf{Alexander Idel}: Conceptualization (supporting); Writing – Review \& Editing (equal). 
\textbf{Jan Peise}: Conceptualization (supporting); Supervision (supporting); Writing – Review \& Editing (equal). 
\textbf{Carsten Klempt}: Conceptualization (lead); Funding Acquisition (lead); Supervision (lead); Writing – Original Draft Preparation (equal); Writing – Review \& Editing (equal). 

\section*{Data Availability}
The data that support the findings of this study are available from the corresponding author upon reasonable request.

\vspace{2cm}\noindent
Copyright (2023) Bernd Meyer-Hoppe, Maximilian Baron, Christophe Cassens, Fabian Anders, Alexander Idel, Jan Peise, Carsten Klempt. This article is distributed under a Creative Commons Attribution (CC BY) License. It may be found at DOI:\href{https://doi.org/10.1063/5.0160367}{10.1063/5.0160367}.

\end{document}